\documentclass[12pt,a4paper]{article}

\usepackage{textcomp}
\usepackage[latin1]{inputenc}
\usepackage{pstricks,pst-node,pst-text}
\usepackage{amsmath}
\usepackage{graphicx}
\usepackage{verbatim}

\newcommand{\bc}{\begin{center}}
\newcommand{\ec}{\end{center}}
\newcommand{\be}{\begin{equation}}
\newcommand{\ee}{\end{equation}}
\newcommand{\bea}{\begin{eqnarray}}
\newcommand{\eea}{\end{eqnarray}}
\newcommand{\beqn}{\begin{eqnarray}}
\newcommand{\eeqn}{\end{eqnarray}}
\newcommand{\ba}{\begin{array}}
\newcommand{\ea}{\end{array}}
\newcommand{\bal}{\begin{aligned}}
\newcommand{\eal}{\end{aligned}}
\newcommand{\ben}{\begin{enumerate}}
\newcommand{\een}{\end{enumerate}}
\newcommand{\bitem}{\begin{itemize}}
\newcommand{\eitem}{\end{itemize}}

\newcommand{\gev}{{\unskip\,\text{GeV}}}
\newcommand{\tev}{{\unskip\,\text{TeV}}}

\newcommand{\fb}{{\unskip\,\text{fb}}}

\newcommand{\fc}{{\texttt{FormCalc-6.0}}}
\newcommand{\fa}{{\texttt{FeynArts}}}
\newcommand{\sloops}{{\texttt{SloopS}}}

\newcommand{\epem}{e^+e^-}

\newcommand{\eezzz}{e^+e^-\to ZZZ}
\newcommand{\eezzzt}{$e^+e^-\to ZZZ\;$}
\newcommand{\eezzztp}{$e^+e^-\to ZZZ$}
\newcommand{\eezzzgam}{e^+e^-\to ZZZ\gamma}
\newcommand{\eewwz}{e^+e^-\to W^+W^- Z}
\newcommand{\eewwzt}{$e^+e^-\to W^+W^- Z \;$}

\newcommand{\eewwzgam}{e^+e^-\to W^+W^- Z\gamma}
\newcommand{\eewwt}{$e^+e^-\to W^+W^-  \;$}

\def\slashepi{\epsilon_i\kern -.720em {/}}
\def\slashpi{p_i\kern -.600em {/}}

\begin{document}

\begin{titlepage}

\vspace*{0.1cm}\rightline{MPP-2010-15}
 \rightline{LAPTH-004/10}

\vspace{1mm}
\begin{center}

{\Large{\bf Electroweak corrections to $W^+W^-Z$ and $ZZZ$ production at the linear collider}}

\vspace{.5cm}

{\bf Fawzi Boudjema, Sun Hao}\\
{\it LAPTH, Universit\'e de Savoie, CNRS,}\\
{\it BP 110, F-74941 Annecy-le-Vieux Cedex, France}

{\bf Le Duc Ninh\footnote{Speaker.}, Marcus M. Weber}\\
{\it Max-Planck-Institut f\"ur Physik (Werner-Heisenberg-Institut),} \\ 
{\it D-80805 M\"unchen, Germany}

\vspace{10mm} \abstract{
We calculate the electroweak corrections to the production of $W^+W^-Z$ and $ZZZ$ 
at the linear collider in the Standard Model. These processes are important for the extraction 
of the quartic couplings of the massive gauge bosons 
which can be a window on the mechanism of spontaneous symmetry breaking. 
We find that the weak corrections to some kinematic distributions 
show new features and hence cannot be explained by an overall scale factor.
}

\end{center}
\vspace*{\fill} {\it The 9th Hellenic School and Workshops on Elementary Particle Physics and Gravity,\\
August 31 - September 6, 2009,\\
Corfu, Greece}


\end{titlepage}

\section{Introduction}
Due to its clean environment an $e^+e^-$ linear collider 
in the $\tev$ range is an ideal machine to probe 
in detail and with precision the inner working of 
the electroweak structure, in particular the mechansim of symmetry breaking.
From this perspective the
study of \eewwzt and \eezzzt may be very instructive and would
play a role similar to $\epem \to W^+W^-$ at lower energies.
Indeed it has been stressed  that  \eewwzt
and \eezzzt  are prime processes for probing the quartic vector
boson couplings \cite{Belanger:1992qh}. In particular deviations from the gauge value in
the quartic $W^+W^-ZZ$ and $ZZZZ$ couplings that are accessible in
these reactions might be the residual effect of physics intimately
related to electroweak symmetry breaking. Since these effects
can be small and subtle, knowing these cross sections with high
precision is mandatory. This calls for theoretical predictions
taking into account loop corrections.

Radiative corrections to \eezzzt have appeared recently in
\cite{JiJuan:2008nn} and those to \eewwzt in \cite{Wei:2009hq}. 
We have made an independent calculation of the
electroweak corrections to \eewwzt and $\eezzz$, see \cite{Boudjema:2009pw}. 
Our preliminary results, eventually confirmed, have been presented in this workshop \cite{ninh-corfu} prior to
\cite{Wei:2009hq}. 
A detailed comparison between our results and the ones of Refs.~\cite{JiJuan:2008nn, Wei:2009hq} 
has been done in \cite{Boudjema:2009pw}. 

In this report we summarize our results and make a further study on some distributions for 
$W^+W^-Z$ production.

\section{Calculational details}
\label{sect-code}
Our calculations are done in the framework of the SM. 
At leading order the process $\eezzz$ contains two types 
of couplings $eeZ$ and $ZZH$. 
This process could probe the effect of a quartic $ZZZZ$ coupling which is absent at tree-level, in the SM. 
The process $\eewwz$ is much more complicated with the involvement of 
trilinear and quartic gauge couplings in addition 
to the similar couplings as in $ZZZ$ production. 
Compared to the well-tested process \eewwt the new ingredients here are the 
two quartic gauge couplings $WWZ\gamma$ and $WWZZ$. Thus, this $WWZ$ production 
at the ILC will be an excellent channel for studying these couplings.  

We have performed the calculation in at least two independent ways
both for the virtual and the real corrections leading to two
independent numerical codes (one code is written in Fortran 77, the other in C++). 
A comparison of both codes has shown
full agreement at the level of the integrated  cross
sections as well as all the distributions that we have studied.\\
\noindent {\bf Input parameters and renormalisation scheme:}\\
We follow closely the on-shell renormalisation scheme
as detailed in Refs.~\cite{grace, Denner:1991kt}. To make the final results 
independent of the light quark masses we adopt a variant of
the $G_\mu$ scheme. At tree level, the electromagnetic coupling constant is
calculated as $\alpha_{G_\mu}=\sqrt{2}G_\mu M_W^2s_W^2/\pi$.
This absorbs some universal $m_t^2$
corrections and also the large logarithmic universal corrections proportional 
to $\ln(q^2/m_f^2)$ where $q$ is some typical energy scale and $m_f$ a fermion mass. 
To avoid double counting we have to subtract the one-loop part of the
universal correction from the explicit ${\cal O}(\alpha)$ corrections
by using the counterterm $\delta Z_e^{G_\mu}=\delta Z_e-\Delta r/2$, 
the expression for $\Delta r$ can be found in \cite{Denner:1991kt}. 
For one-loop corrections we use the coupling $\alpha(0)$ for both virtual
and real photons. Thus the NLO corrections are of order
$\alpha_{G_\mu}^3\alpha(0)$. Further details and the complete set of input parameters 
are given in \cite{Boudjema:2009pw}.\\
\noindent {\bf Virtual corrections:}\\
The virtual corrections have been evaluated using a conventional
Feynman-diagram based approach using standard techniques. 
We use the packages \fa\ and \fc\ to generate all Feynman diagrams 
and helicity amplitude expressions \cite{fafc}. 
We also use \sloops\ to check the 
correctness of the amplitudes by checking non-linear gauge invariance 
(see \cite{sloops} and references therein). 
The total number of diagrams in the 't~Hooft-Feynman
gauge is about 2700
including 109 pentagon diagrams for $\eewwz$ and about 1800
including 64 pentagons for $\eezzz$. This already shows that
$\eewwz$ with as many as 109 pentagons is more challenging than
$\eezzz$. 
Indeed getting stable results for all scalar and tensor (up to rank 4) box integrals in 
the process $\eewwz$ is a highly nontrivial task. 
An efficient way to solve this problem is by using higher precision arithmetic in 
part of the calculation. Further details related to loop integrals and 
the references for useful public codes are given in \cite{Boudjema:2009pw}.\\
\noindent {\bf Real corrections:}\\
In addition to the virtual corrections we also have to consider
real photon emission, {\it i.e.} the processes $\eewwzgam$ and
$\eezzzgam$. The corresponding amplitudes are divergent in the
soft and collinear limits. The soft singularities cancel against
the ones in the virtual corrections while the collinear
singularities are regularized by the physical electron mass.  To
extract the singularities from the real corrections and combine
them with the virtual contribution we apply both the dipole
subtraction scheme and a phase space slicing method. 
The former is used to produce the final results since it yields smaller 
integration errors. Further details are given in \cite{Boudjema:2009pw}. \\
\noindent {\bf Defining the  weak corrections:}\\
It is well-known that the collinear QED correction related to
initial state radiation in $e^{+}e^{-}$-processes is large.
In order to see the effect of
the weak corrections, one should separate this large QED correction from
the full result. It means that we can define the weak correction as
an infrared and collinear finite quantity. The definition we adopt
in this paper is based on the dipole subtraction formalism. In this
approach, the sum of the virtual and the so-called "endpoint" 
(see \cite{Dittmaier:1999mb} for the definition) contributions satisfies
the above conditions and can be chosen as a definition for the weak
correction
\beqn
\sigma_\text{weak} = \sigma_\text{virt} + \sigma_\text{endpoint}.
\eeqn
For the numerical results shown in the next section,
we will make use of this definition.
\section{Numerical results}
\label{sect-results}
\begin{figure}[t]
\begin{center}
\mbox{\includegraphics[width=0.47\textwidth]{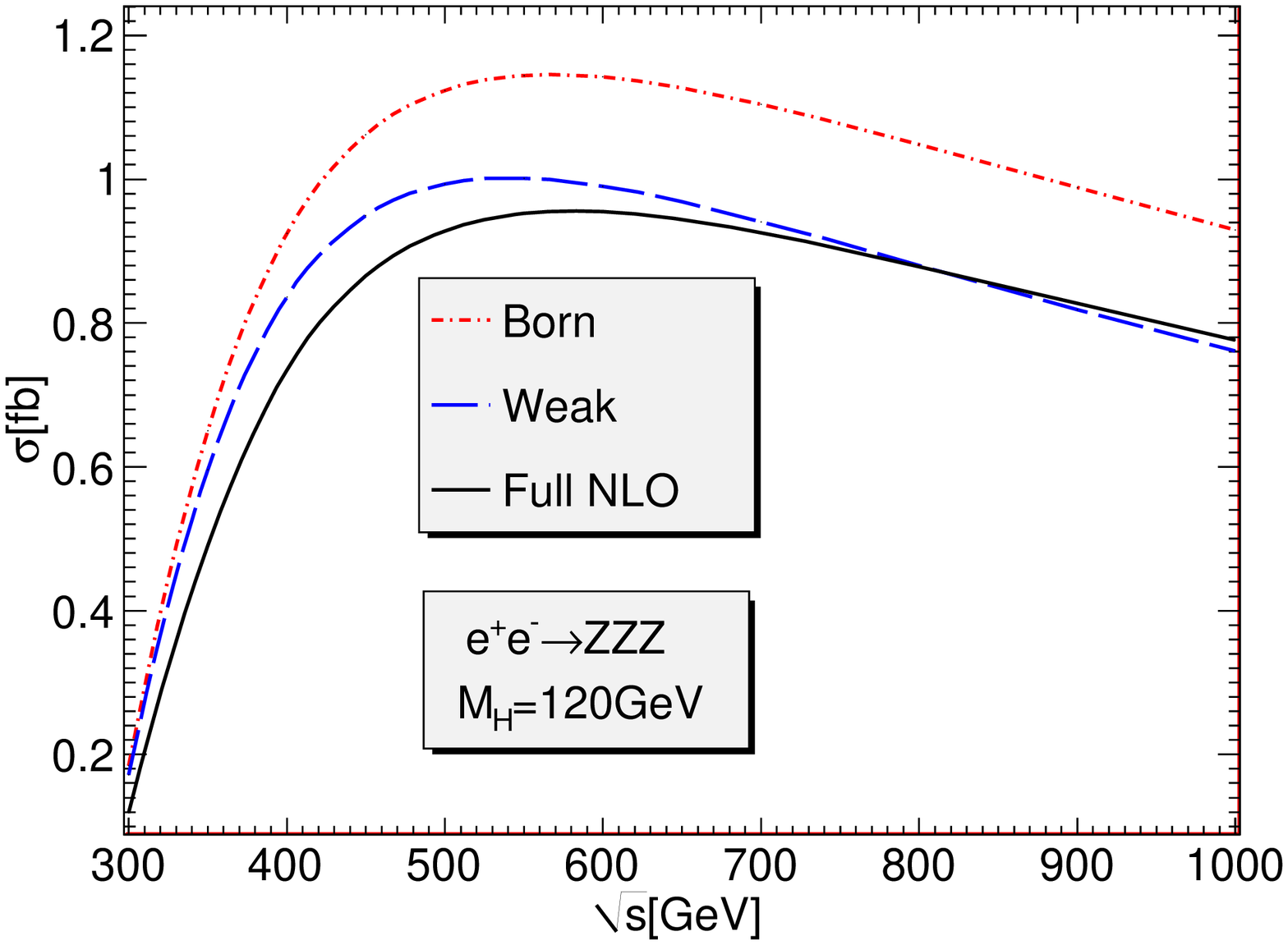}
\hspace*{0.01\textwidth}
\includegraphics[width=0.47\textwidth]{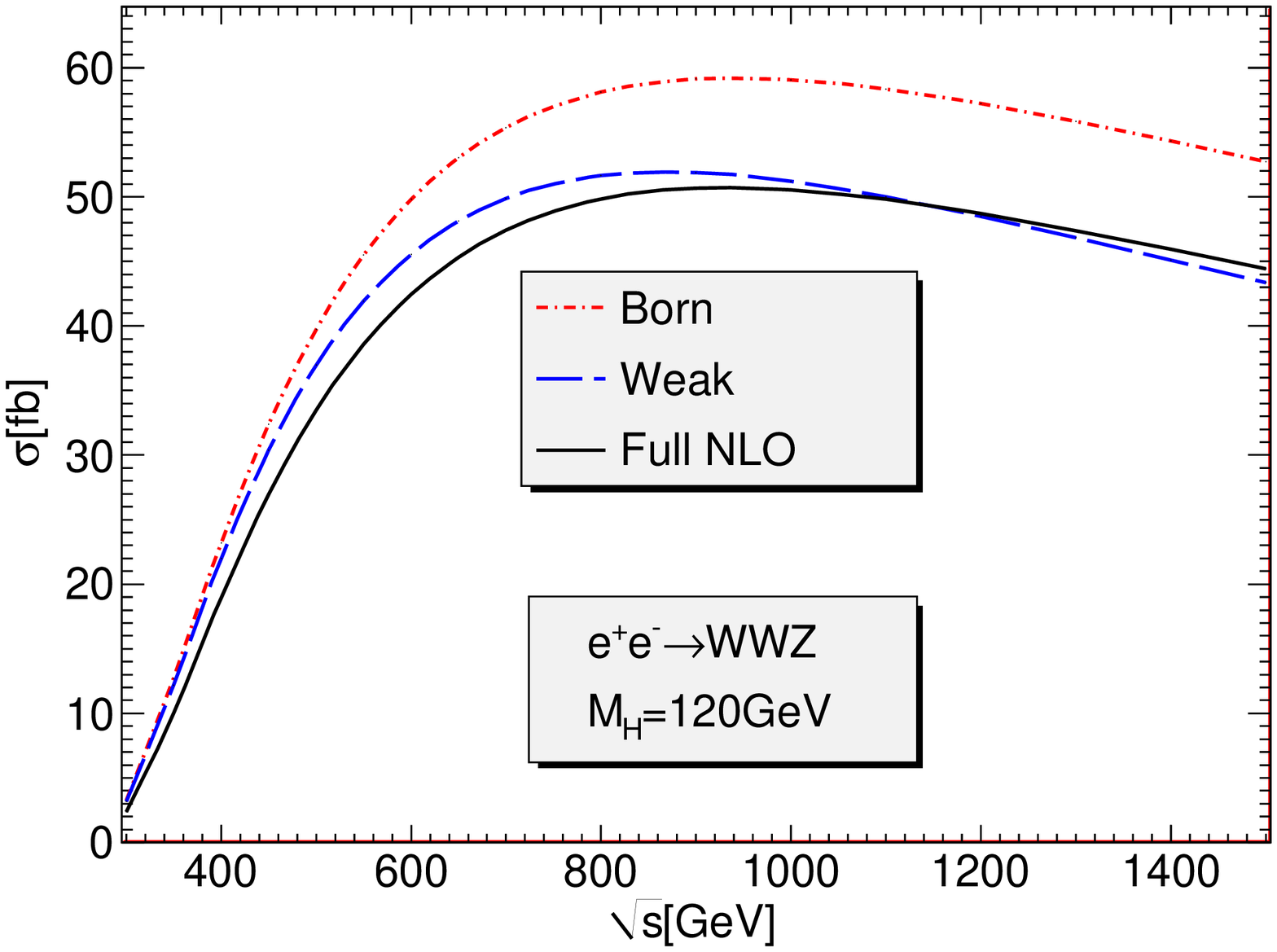}}
\caption{\label{ourvvzsigma}{\em Left: the total cross section
for $\eezzz$ as a function of $\sqrt{s}$ for the Born, full ${\cal
O}(\alpha)$
 and genuine weak correction. Right: the
same for $\eewwz$.}}
\end{center}
\end{figure}
\noindent {\underline{$\eezzz$:}}\\
As shown in Fig.~\ref{ourvvzsigma} the tree-level cross section
rises sharply once the threshold for production opens, reaches a
peak of about $1.1\fb$ around a centre-of-mass energy of $600\gev$
before very slowly decreasing with a value of about $0.9\fb$ at
1\tev. The full NLO
corrections are quite
large and negative around threshold, $-35\%$, decreasing sharply to stabilise
at a plateau around $\sqrt{s}=600\gev$ with $-16\%$ correction. The
sharp rise and negative corrections at low energies are easily
understood. They are essentially due to initial state radiation
(ISR) and the behaviour of the tree-level cross section. The
photon radiation reduces the effective
centre-of-mass energy and therefore explains what is observed in the
figure.
On the other hand the genuine weak corrections, in the
$G_\mu$ scheme, are relatively small at threshold, $-7\%$. They
however  increase steadily with a correction as large as $-18\%$
at $\sqrt{s}=1\tev$.\\
\noindent {\underline{$\eewwz$:}}\\
Compared to $ZZZ$ production, the cross section for
\eewwzt is almost 2 orders of magnitudes larger for the same
centre-of-mass energy. For example at $500\gev$ it is about $40\fb$ at tree
level,
compared to $1\fb$ for the \eezzzt cross section. For an anticipated
 luminosity of $1 {\rm ab}^{-1}$, this means that
the cross section should be known at the
per-mil level. The behaviour of the total cross section as a function of
energy resembles that of \eezzztp. It rises sharply once the
threshold for production opens, reaches a peak before very slowly
decreasing as shown in Fig.~\ref{ourvvzsigma}. However as already
discussed the value of the peak is much larger, $\sim 50\fb$ at NLO,
moreover the peak is reached around $\sqrt{s}=1\tev$, much higher than
in $ZZZ$. This explains the bulk of the NLO corrections at lower
energies which are dominated by the QED correction, large and
negative around threshold and smaller at higher energies. As
the energy increases the weak corrections get larger
reaching about $-18\%$ at $\sqrt{s}=1.5\tev$. 
\begin{figure}[htb]
\begin{center}
\mbox{\includegraphics[width=0.45\textwidth
]{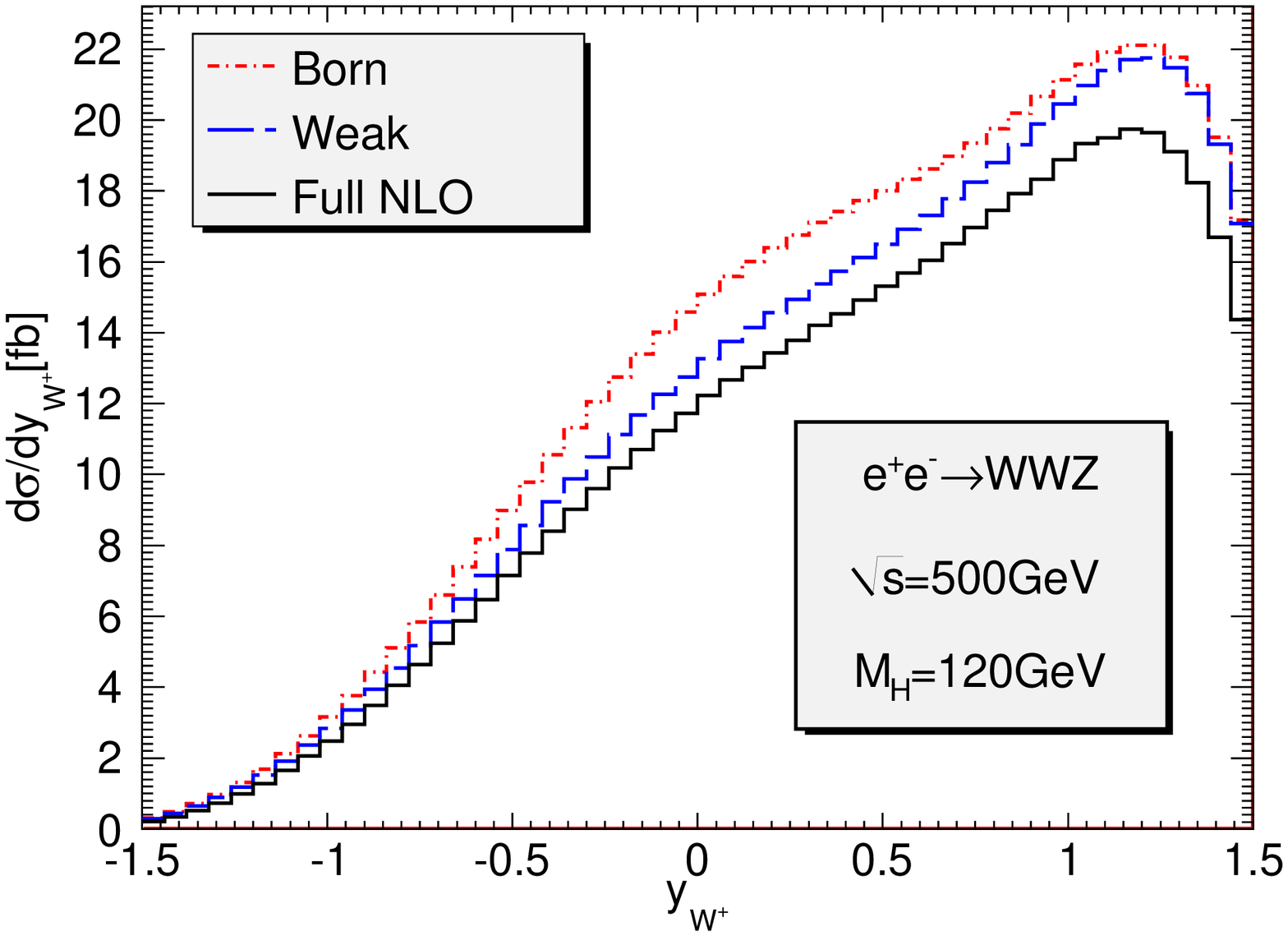} \hspace*{0.01\textwidth}
\includegraphics[width=0.45\textwidth]{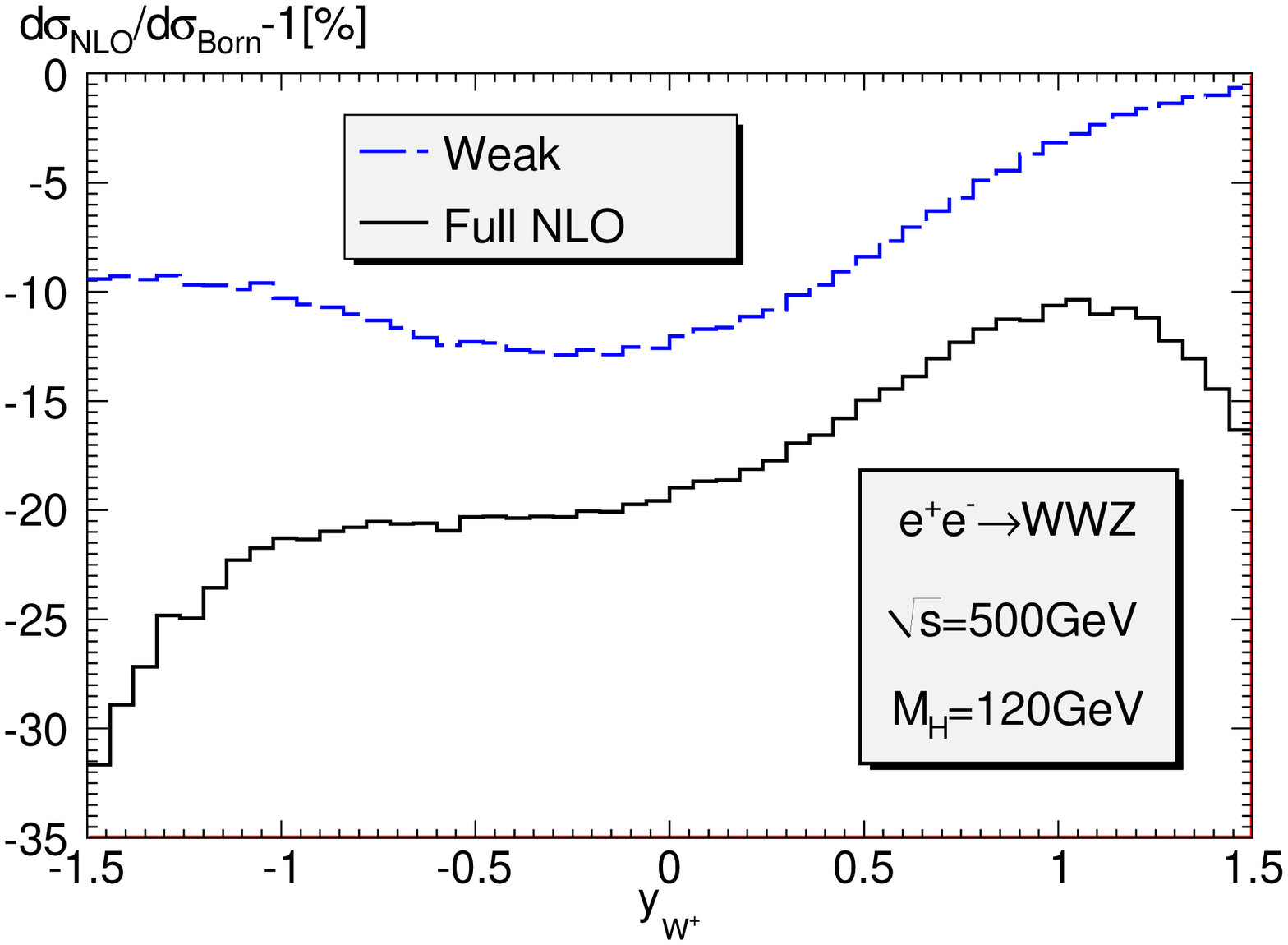}}
\mbox{\includegraphics[width=0.45\textwidth
]{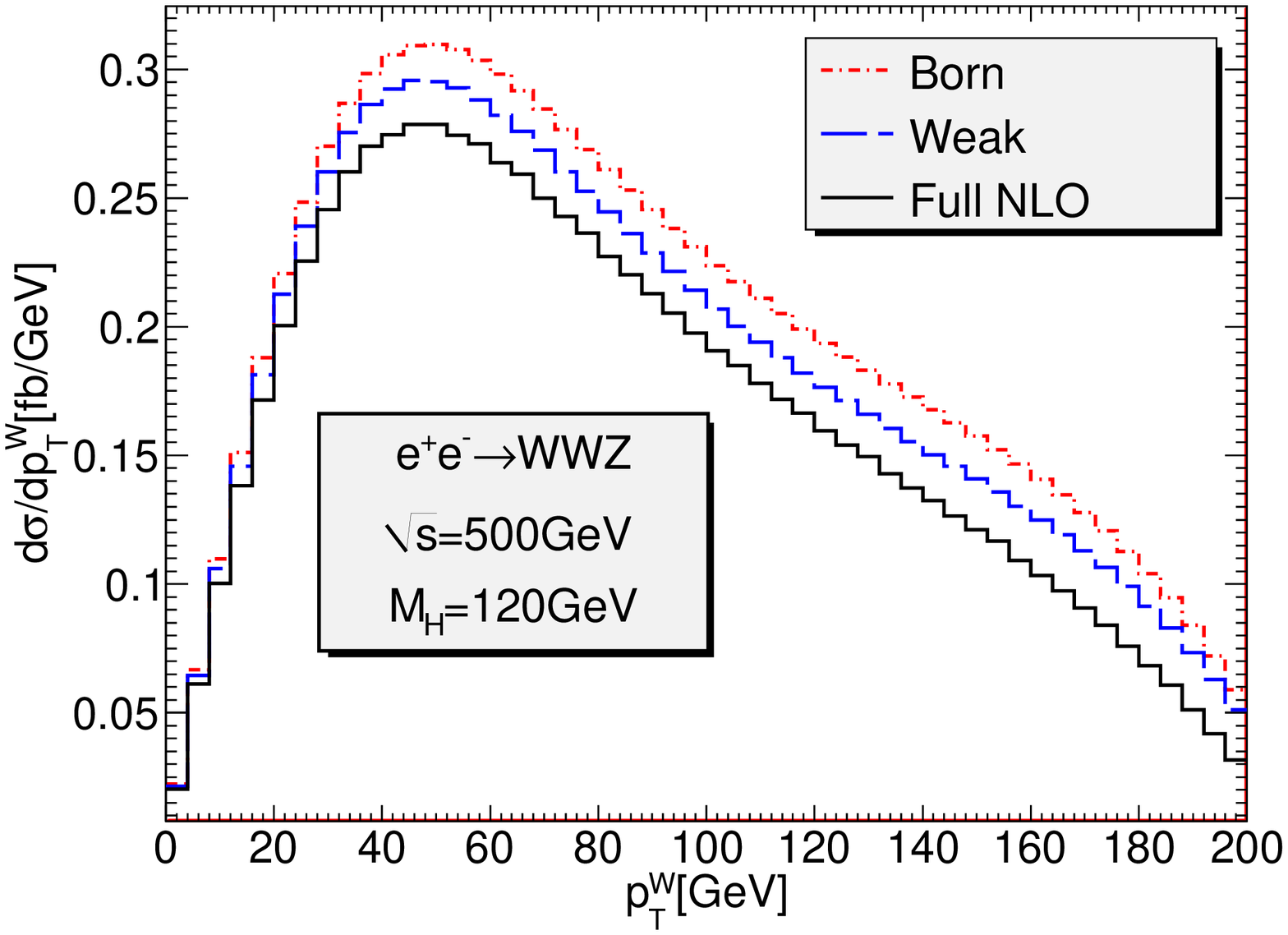} \hspace*{0.01\textwidth}
\includegraphics[width=0.45\textwidth]{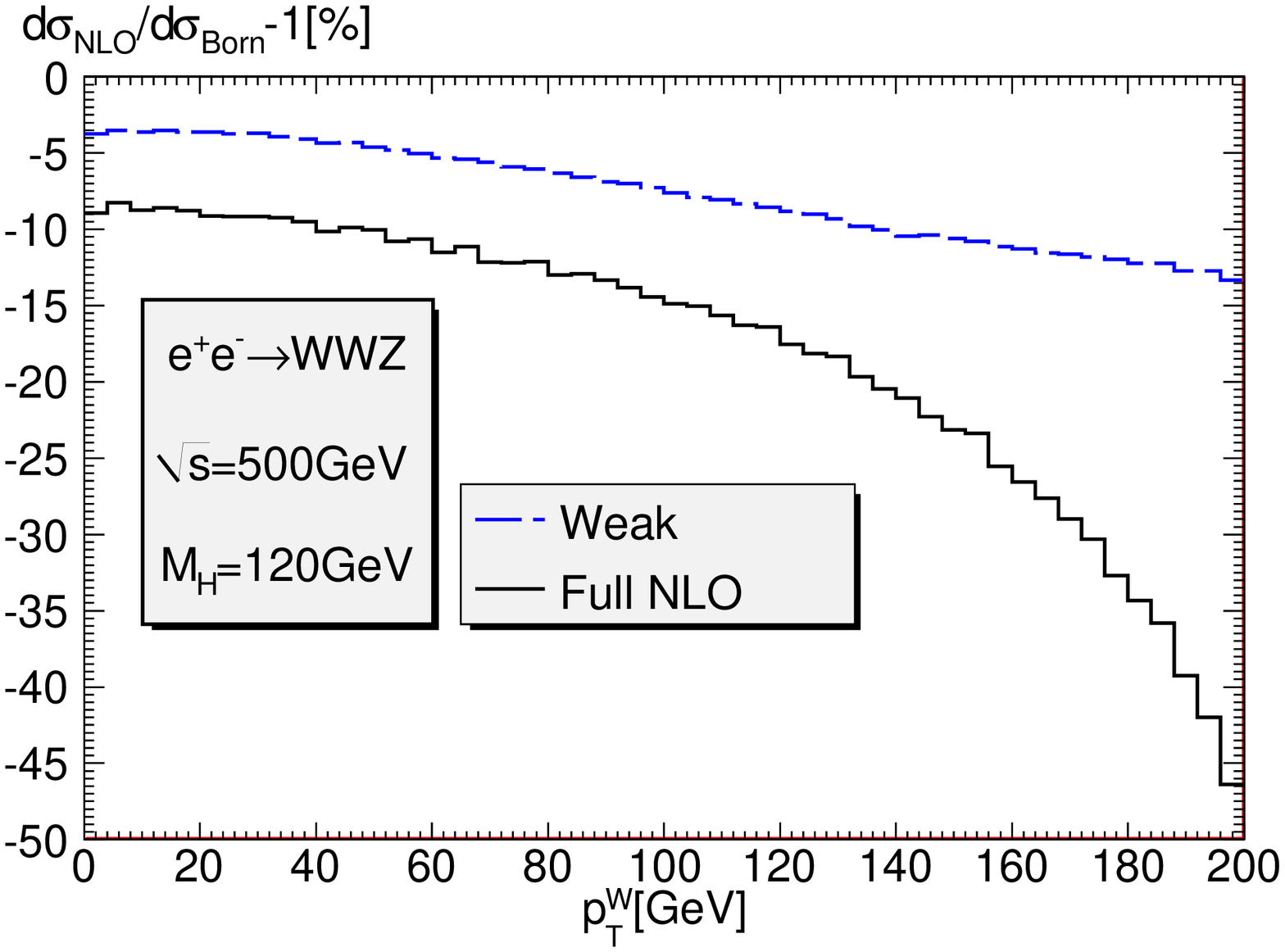}}
\caption{\label{ourwwzdist}{\em From top to bottom: distributions
for the rapidity and the transverse momentum of the $W^+$ for \eewwzt . 
The panels on
the left show the tree-level, the full NLO and the weak correction.
The panels on the right show the
corresponding relative (to the tree-level) percentage corrections.
}}
\end{center}
\end{figure}

In Fig.~\ref{ourwwzdist} we show the distributions in 
the rapidity and the transverse momentum of the $W^+$. 
First, due to photon
radiation, in the full NLO corrections some large corrections do
show up at the edges of phase space.
However when the QED corrections are subtracted, the weak
corrections cannot be parameterized by an
overall scale factor, for all the distributions that we have
studied. This feature together with possible normal/anomalous thresholds make it clear that 
calculating explicitly one-loop EW corrections is needed for a precise comparison with experimental 
data.



\end{document}